\documentclass{article}
\usepackage{arxiv}
\usepackage[utf8]{inputenc} 
\usepackage[T1]{fontenc}    
\usepackage{hyperref}       
\usepackage{url}            
\usepackage{booktabs}       
\usepackage{nicefrac}       
\usepackage{microtype}      
\usepackage{amsmath,amsfonts,amssymb,amsthm,mathtools,graphicx,colortbl} 
\usepackage{subfig,multirow,rotating}
\usepackage{pdflscape,longtable,threeparttablex}

\title{Targeted Estimation of Heterogeneous Treatment Effect in Observational Survival Analysis}

\author{
  Jie Zhu \\
  Centre for Big Data Research in Health (CBDRH)\\
  University of New South Wales (UNSW)\\
  \texttt{elliott.zhu@student.unsw.edu.au} \\
   \And
  Blanca Gallego \\
  Centre for Big Data Research in Health (CBDRH)\\
  University of New South Wales (UNSW)\\
}

\begin{document}
\maketitle

\begin{abstract}
The aim of clinical effectiveness research using repositories of electronic health records is to identify what health interventions 'work best' in real-world settings. Since there are several reasons why the net benefit of intervention may differ across patients, current comparative effectiveness literature focuses on investigating heterogeneous treatment effect and predicting whether an individual might benefit from an intervention. The majority of this literature has concentrated on the estimation of the effect of treatment on binary outcomes. However, many medical interventions are evaluated in terms of their  effect on future events, which are subject to loss to follow-up. In this study, we describe a framework for the estimation of heterogeneous treatment effect in terms of differences in time-to-event (survival) probabilities. We divide the problem into three phases: (1) estimation of treatment effect conditioned on unique sets of the covariate vector; (2) identification of features important for heterogeneity using an ensemble of non-parametric variable importance methods; and (3) estimation of treatment effect on the reference classes defined by the previously selected features, using one-step Targeted Maximum Likelihood Estimation. We conducted a series of simulation studies and found that this method performs well when either sample size or event rate is high enough and the number of covariates contributing to the effect heterogeneity is moderate.  An application of this method to a clinical case study was conducted by estimating the effect of oral anticoagulants on newly diagnosed non-valvular atrial fibrillation patients using data from the UK Clinical Practice Research Datalink.
\end{abstract}

\keywords{Survival Analysis \and Machine Learning \and Heterogeneous Treatment Effect \and Targeted Maximum Likelihood Estimation}

\section{Introduction}\label{intro}

	In medical studies, we want to use empirical evidence to estimate the effect of treatment: such as a drug or a procedure. The gold standard for the evaluation of treatment effect is the randomized control trial (RCT) since its randomisation minimises bias and maximises our ability to identify causality. However, it has become clear that it is not possible to depend on RCTs for all the information needed on the effectiveness of medical interventions since they do not represent real-world populations or settings and tend to be too short to detect long-term effects. Furthermore, RCTs are designed to estimate the average effect of interventions and might, therefore, not be able to inform decisions about individual patients encountered in clinical practice. Recently, there has been an explosion of observational studies where large repositories of routinely collected health data are used to customize estimates of treatment effect for individuals. In this study, we focus on the estimation of heterogeneous treatment effect (HTE) for binary treatment options (treated vs control) on time-to-event (survival) outcomes using right-censored data. These estimates account for loss to follow-up and allow for the measurement of the long-term effect of interventions.

    Commonly used measures of treatment effect in terms of survival outcomes are the hazard ratio, the difference in length of survival time, and the difference in survival probability, of which only the latter two are absolute effect measures. Current standards for the reporting of findings in experimental studies (CONSORT\cite{Schulz:2010gw}) and observational studies (STROBE\cite{vonElm:2007hb}) recommend reporting both absolute and relative measures of effect whenever possible. In spite of this recommendation, the majority of previous clinical survival analyses report hazard ratios under Cox proportionality assumptions \cite{Cox:1972cw}. This is not ideal for the estimation of HTE since absolute effects provide more relevant information for clinical decisions and should be used when describing personalised treatment effects \cite{Kent:2018ii}. Furthermore, the proportionality of hazards assumption does not always hold. Here, we choose the difference in survival probabilities as the primary treatment effect measure and convert it into a hazard ratio when required for comparison with other studies.

    The estimation of HTE in survival analysis requires us to simultaneously adjust for selection bias from both censoring and treatment allocation and find the reference class associated with an individual patient for which reliable statistics can be compiled. That is, our goal is to estimate a conditional average treatment effect (CATE) for all heterogeneous reference classes. Methods for HTE estimation generally fall into one of three categories: transformed outcome regression \cite{Jaskowski:2012ub}, conditional mean regression \cite{Powers:2018bw,Wendling:2018cu}, and double-robust estimators \cite{2019arXiv190105056B}. 

    The transformed outcome approach creates a new target variable that combines the observed outcomes weighted by the inverse of the probability that a patient would receive treatment given patient covariates (propensity score). It has the property that under the assumption of unconfoundness\cite{Jaskowski:2012ub} \cite{Rosenbaum:1983ct}, the estimation of CATE for each unique set of covariates is unbiased, although at the cost of high variance. 

    Conditional mean regression models aim at accurately estimating condition mean functions for the potential outcomes and deriving treatment effect as the difference between these potential outcomes. These methods can suffer from bias if the outcome model is misspecified, particularly in areas of the covariate space where there is no overlap between intervention and control. Recent applications of conditional mean regression models in survival analysis report the effect either as the difference in length of survival time using random forest  \cite{Tabib:vz,Zhang:vx}; or in terms of survival probability using a deep neural network with Cox proportionality of hazards assumption (DeepSurv\cite{Katzman:2018bv}), a deep neural network accepting competing risks (DeepHit\cite{Lee:2018tc}), a deep recurrent neural network (DRSA\cite{Ren:2018ue}), or Super-learner \cite{Polley:2011jw}.

    Double-robust estimators refer to a class of procedures that combine outcome models with propensity score models. In the survival context, they are double-robust in the sense that the treatment effect estimator is consistent if either the propensity score and the conditional probability of censoring are consistently estimated, or the conditional survival probability of failure event is consistently estimated. These estimators enjoy flexible regression tools such as those used in conditional mean regression models and are often statistically more efficient (unbiased) due to the use of propensity scores. Existing frameworks for efficient causal effect estimation include the targeted maximum likelihood estimation (TMLE)\cite{vanderLaan:2011ep}, the Quasi-Oracle R-learner\cite{2017arXiv171204912N}, and the recent Non-stationary Gaussian Process Regression (NSGP) \cite{Alaa:2018ue} which assumes a non-stationary Gaussian process prior as the initial estimator instead of the conditional mean in TMLE. Although most methods developed in this class are aimed at uncensored outcomes, it is possible to use TMLE to estimate average survival probability at a single time point \cite{benkeser2017improved}. Furthermore, recent development of one-step TMLE \cite{Cai:2018vt} enables the estimation of a monotonic average treatment effect curve. 

    Investigating the performance of double-robust estimators in survival causal inference motivates this paper. To estimate HTE as the difference in survival probabilities, we face three challenges: 1) model the unobserved survival probability using longitudinal data; 2) identify key covariates contributing to the heterogeneity of the survival treatment effect; and 3) form double-robust estimates of CATE. We address the first challenge by converting the survival dataset into a counting process (i.e., longitudinal data series with binary outcomes) and computing the cumulative hazard rate to get the survival probabilities. This allows us to circumvent the Cox proportionality assumption as well as to use data-adaptive outcome models such as Super-learner or Deep learning techniques. In this study, we choose Super-learner algorithms on moderate-dimensional data and leave high dimensional deep learning techniques for future work. To address the second challenge, we use an ensemble of standard methods used to identify variable importance in causal inference, including Lasso\cite{imai_strauss_2011}, Causal Forest\cite{Wager:2018eq} and permutation test using BART\cite{bartMachine}. The Kneedle Algorithm\cite{Satopaa:kneedle} is applied to each method to provide the threshold for variable inclusion. Finally, CATE is estimated for the reference classes defined by the selected features using kernel-based local averaging \cite{Abrevaya:2014fk}, and selection bias is adjusted for using one-step TMLE. 

    An outline of this paper is as follows. Section 2 describes the methodology to estimate the survival treatment effect. In Section 3, we introduce a set of simulation scenarios under different data dimensions, sizes, confounding levels, and event rates. Sections 4 provides a case study that estimates the effect of anticoagulants in atrial fibrillation patients. We end with a discussion.

\section{Targeted survival heterogeneous effect estimation}\label{sec2}

	\subsection{Definition of survival treatment effect}
		To formalize the framework for causal inference in survival analysis, we follow the notations in previous studies \cite{vanderLaan:2011ep,Imai:2013ho,Abrevaya:2015hw}. Suppose we observe a sample $\mathcal{O}$ of $n$ individuals generated by an unknown distribution $\mathcal{P}_0$:
		\begin{eqnarray*}
			\mathcal{O} \coloneqq (X_i,A_i,Y_i, T_{i}=\min(t_{s,i},t_{c,i})),  i = 1,2,\ldots, n,
		\label{e0}
		\end{eqnarray*}
		where $X_i = (X_{i,1},X_{i,2},\ldots,X_{i,\textit{j}}), j=1,2,\ldots,J$ are baseline covariates; $A_i$ is the exposure condition, $A_i=1$ if observation \textit{i} receives the treatment and $A_i=0$ otherwise; $Y_i(t)$ denotes the outcome at time $t$, $Y_i=1$ if \textit{i} experienced an event and $Y_i=0$ otherwise; $T_{i}$ is the observed event/censoring time and is determined by the potential event occurrence time, $t_{s,i}$, or censor time $t_{c,i}$. We can define the conditional survival function $S(t)_i$ at time $t$ for $i$ as:
		\begin{eqnarray*}
		S(t_i = t|A_i,X_i) \coloneqq P(t_i >t |A_i,X_i),  t = 1,2,\ldots,\Theta,
		\end{eqnarray*}
		\noindent where $\Theta$ is the end of follow up period. Using the potential outcomes framework by Rosenbaum and Rubin \cite{Rosenbaum:1983ct}, we define unconfoundness by assuming: 1) the treatment assignment $A$ is independent of event/censoring time $T$ given $X$; and 2) the censoring is non-informative (coarsening at random). Under these assumptions, the conditional average treatment effect (CATE), $\psi(x,t)$, can be defined as:
		\begin{eqnarray}
		\psi(X,t) &= \mathbb{E}_{X=x}\big[\mathbb{E}[S(t)|A=1,X=x]-\mathbb{E}[S(t)|A=0,X=x]\big].
		\label{e:cate}
		\end{eqnarray}
		A reliable estimation of CATE is straightforward when there are large amount of observations for each unique set of covariates \cite{Grimmer:2017ga}  $X=x$. However, this is unlikely in common medical studies. For a single observation with $X=x$, equation \eqref{e:cate} represents an individual treatment effect (ITE). For the purpose of this manuscript, $\psi(X,t)$ is called ITE when x represents the full covariate vector. The aim of this study is to identify important features $X$ contributing to the heterogeneity of $\psi(X,t)$, and form double-robust estimations of CATE for these selected reference classes.

		\begin{figure}[htbp!]
		
		\centerline{\includegraphics[height=8pc]{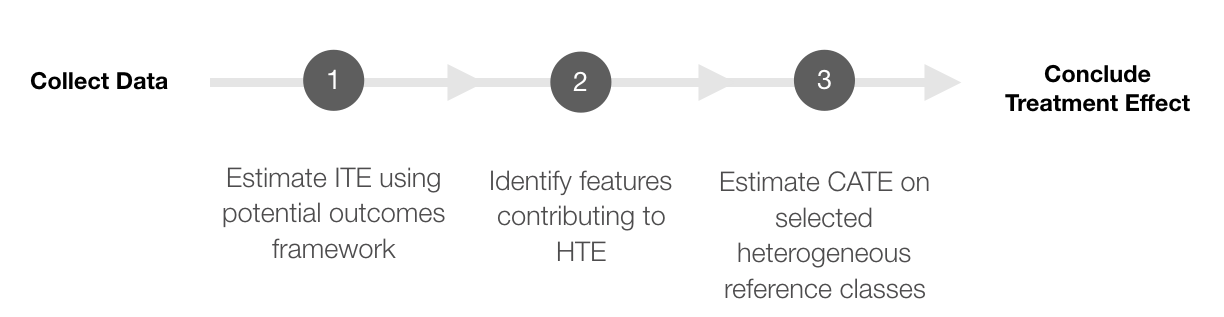}}
		\caption{\fontsize{8pt}{10pt}\selectfont Diagram depicting heterogeneous effect estimation framework.\label{fig1}}
		\end{figure}

	\subsection{The framework}

		Figure \ref{fig1} depicts the proposed framework for CATE estimation, we describe the three steps in the sections below. 

		\paragraph{Step 1. Estimate of individual treatment effect}

		Individual treatment effects are estimated by fitting outcome models on treated and control samples: $f(t|A=1,X)$ and $f(t|A=0,X)$, $t =1,2,\ldots,\Theta$; then, we calculate potential outcomes using these models as $\mathbb{E}[S(t)_i]=f(t|A=1,x_i)$ and $\mathbb{E}[S(t)_i]=f(t|A=0,x_i)$. This two-model approach encodes the different relevant explanatory variables for treated and control outcomes and has been proved to improve the potential outcome estimation accuracy \cite{Alaa:2018ue}. 
		
		We start by converting the original survival data into a counting process: a sequence of binary outcomes over time. We then use Super-learner to estimate the conditional hazard rate $h(t)_i$ for each observation $i$ at each time $t$ as: 
		\begin{eqnarray*}
		h(t)_i \coloneqq P(t_i \in (t-1,t] |t_i >t-1,A_i,X_i),
		\end{eqnarray*}
		where $t_i$ is the observed event time for $i$. The potential survival probabilities are then derived via the \textit{probability chain rule}: 
		\begin{eqnarray*}
		S(t)_i = \prod_{T:T\leq t} (1-h(T)_i), 
		\end{eqnarray*}
		and plug into equation \eqref{e:cate} to get the initial treatment effect estimation $\hat\psi(X_i,t)$ for $i$ at time $t$. 
		
		The accuracy of this estimation at each time $t$ is computed using normalized root-mean squared error (NRMSE):

		\begin{eqnarray*}
		  \mathrm{NRMSE(t)} = \frac{1}{\bar \psi(t)} \sqrt{ \sum_{i=1}^{n}\big(\hat \psi(X_i,t) -  \psi(X_i,t)\big)^2},   t =1,2,\ldots,\Theta, 
		\end{eqnarray*}
		\noindent where $\bar \psi(t)$ is the estimated average treatment effect at time $t$ and  $\psi(X_i,t)$ represents the unobserved true individual treatment effect.

		\paragraph{Step 2. Identification of features contributing to heterogeneity of treatment effect}

		Given the estimated ITE, we seek to identify which features contribute to the heterogeneity of treatment effect. We model ITE using  machine learning algorithms and define important features as those that contribute most to the reduction of the model prediction error. The regression methods employed here are: Bayesian Additive Regression Trees (BART)\cite{bartMachine}, Adaptive Lasso (AL)\cite{Zou:2006du}, Elastic Net (EN)\cite{Zou:2005fc} and Causal Forest (CF)\cite{Wager:2018eq}. Previous studies found AL and EN give consistent selections of covariates compared to the standard Lasso.\cite{Zou:2006du,Zou:2005fc} Alternatively, we could have chosen to model ITE within strata of similar propensity score\cite{Xie:2012wy}. However, this was not chosen since recent work has demonstrated the propensity score matching (PSM) may result in increased covariate imbalance in observational studies.\cite{226731}

		The feature identification procedure is as follows:

		\begin{enumerate}
			\item Regress the estimated ITEs at the end of follow-up period $\Theta$ on covariates $X$ using model $f(\cdot)$: 
			\begin{eqnarray*}
			 \hat \psi(X,\Theta) = f(X);
			\end{eqnarray*}

		   \item Calculate the variable importance score $S_j$ for features $X_j, j=1,2,\ldots,D$, where $D$ is the number of features;

		   \item Rank features according to their importance score $S_j$ in descending order and assign the rank $R_j$ to each $X_j$;

		   \item Construct an importance score curve using ordered pairs $(S_j,R_j)$; 

		   \item Use \textit{Kneedle Algorithm}\cite{Satopaa:kneedle} to identify the \textit{Knee Point}\footnote{One can think \textit{Knee Point} as the inflection point at the local minimum (if the importance score curve is convex) or the local maximum (if the curve is concave).  Please refer to Appendix A for more detail.}, and label features with a higher importance score than the \textit{Knee Point} as having significant contribution to the heterogeneous effect.   
		\end{enumerate}

		\paragraph{Step 3. Target estimation of CATE}

		 Kernel-based local averaging\cite{Abrevaya:2014fk} is used to calculate CATE associated with the reference classes defined by the previously selected features. Let $X^*$ be a selected feature, we can divide the population into $Q$ strata ($q=1,2,\ldots,Q$) based on the value of $X^*$, then for each stratum $q$ ($q=1,2,\ldots,Q$), we  calculate the marginal conditional average treatment effect (MCATE) as:
		\begin{eqnarray*}
		    \psi_q(X^*,t) = \frac{1}{h_q}\sum_{i=1}^{h_q} \psi^*(X_i,t|_{X^*\in q}),   t =1,2,\ldots,\Theta, 
		\end{eqnarray*}
		\noindent where $h_q$ is the number of observations in stratum $q$ and $\psi^*(\cdot)$ is the one-step TMLE adjusted treatment effect. To get $\psi^*(\cdot)$, we adjust the initial estimator $\hat\psi(\cdot)$ in step one with conditional censoring probability, $P(C>t |A,X)$, and the propensity score, $P(A=1|X)$, where $t = 1,2,\ldots,\Theta$ and $C$ is the observed censor time. We estimate both probability scores and conduct the adjustment following the previous study\cite{Cai:2018vt}. The accuracy of CATE estimations is measured by absolute percentage bias:

		\begin{eqnarray*}
		  \%\mathrm{Bias(t)} = \bigg|  \frac{\hat \psi_q(X^*,t) -  \psi_q(X^*,t) }{\hat \psi_q(X^*,t)} \bigg|  ,  \text{  } t =1,2,\ldots,\Theta, \text{ and, } q = 1,2,\ldots,Q.
		\end{eqnarray*}

		All computations were carried out using R software. For ITE, censoring probability and propensity score estimations, we use R package \textit{SuperLearner}\cite{SuperLearner}, with an ensemble of generalized additive models (R package \textit{gam}\cite{gam} at degrees of freedom at $0.5$), multivariate adaptive regression splines (R package \textit{earth}\cite{earth}) and elastic-net regularized generalized linear models (R package \textit{glmnet}\cite{glmnet}). The training of Super-learner is cross-validated with 10 folds. To calculate variable importance, we use R package \textit{grf}\cite{grf} for causal forest, \textit{BartMachine}\cite{bartMachine} for BART, and \textit{glmnet}\cite{glmnet} for Adaptive Lasso and Elastic Net. For one-step TMLE adjustment, we use package \textit{MOSS}\cite{MOSS}.

\section{Experiments}

\subsection{Design of experiments}
		To explore the finite-sample performance of our proposed method, we conduct simulations under multiple scenarios. Survival datasets are generated following previous literatures \cite{Cai:2018vt,vanderLaan:2018bf,PeterCAustin:2015gv} using:

		\begin{itemize}
			\item $D-1$ continuous covariates $X_1,X_2, X_3, X_5,X_6,\ldots,X_D \sim  \mathrm{Unif}(0,1)$;
			\item a binary covariate $X_4 \sim \mathrm{Binom}(0.5)$;
			\item a binary exposure $A \sim \mathrm{Binom}( \mathrm{odds} / (1 + \mathrm{odds}))$, where $\mathrm{odds} = 0.25+\beta(X_1+X_2)$ and  $\beta$ controls the level of confounding; 
			\item survival time $T \sim  \mathrm{exp} (\tau(A,X))$, where 
			\begin{eqnarray*}
				\tau(A,X) = (A(X_1+3X_5+(1-3X_2)^2+X_3X_4)+X_1+X_2+\ldots+X_D)/r;
			\end{eqnarray*}
			The component $X_1+3X_5+(1-3X_2)^2+X_3X_4$ controls the heterogeneous effect. It combines two linear features of different scale, $X_1$ and $3X_5$, a non-linear feature $(1-3X_2)^2$, and an interaction term $X_3X_4$. The denominator $r$ controls the magnitude of $\tau(A,X)$, a larger $r$ leads to lower event probability and longer survival time. $r$ also controls the event rate $\mathcal{R} = \text{number of events / number of observations}$. In this design, an observation with treatment condition $A=1$ will have ceteris paribus lower survival probability than $A=0$;
	 		\item censor time $C\sim Weibull(1+0.2X_1, 50)$, which is lightly confounded by $X_1$; and
	 	    \item survival outcome $Y=I(T\leq C)$, where $I$ is an indicator function, $I=1$ if $T\leq C$ and $I=0$ otherwise. The outcome is confounded on covariates $X_1$ and $X_2$. Thus, the true survival probability at $t$ is $S(t|A,X) = \mathrm{exp}(-t \tau(A,X))$. 

		\end{itemize}
	
		A series of experiments are conducted by changing the following simulation parameters:

		\begin{itemize}
			\item Sample size, $N$ (default: $N=3,000$; range: $N=1,000, 3,000 , 9,000$ and $15,000$);
			\item Dimension, $D$ (default: $D=10$; range: $D=8,10,14,20,30$);
			\item Level of confounding, $\beta$ (default: $\beta=0.5$; range: $\beta=0, 0.5, 1.0, 2.0$);
			\item Event rate, $\mathcal{R}$ (default:  $\mathcal{R}=10\%$; range: $ \mathcal{R}=2.5\%, 5.0\%, 10\%, 15\%,20\%,30\%$); and
			\item Number of subgroups, $M$ (default: $M=10$; range: $1, 5, 15,\ldots, 50$). When $M=1$, we make estimations using the whole sample and get the average treatment effect (ATE); when $M>1$, the subgroup size for a given continuous covariate will roughly equals $N/M$ since we simulate continuous covariates using the uniform distribution. 
		\end{itemize}
	
		The default values have been chosen as typical values encounter in medical studies and previous literature. 

		In the first set of experiment, we examine the accuracy of feature identification under different values of $N, D, \beta$ and $\mathcal{R}$. The impact of changing $N$ and $D$ is examined with default values $\beta=0.5$ and $\mathcal{R}=10\% $. The impact of changing $\beta$ and $\mathcal{R}$ is examined with default values $N=3,000$ and $D=10$. Two measures are used for benchmarking: 1) positive predictive value ($\mathrm{PPV}=\mathrm{TP}/(\mathrm{TP}+\mathrm{FP})$) and 2) true positive rate ($\mathrm{TPR} = \mathrm{TP}/\mathrm{P}$), where $\mathrm{P}$ is the number of true features explaining the heterogeneous effect, $\mathrm{TP}$ is the number of correctly predicted features, and $\mathrm{FP}$ is the number of falsely predicted features. The second experiment examines the accuracy of CATE estimations under different values of $N, D, \beta, \mathcal{R}$ and $M$.

      %

\subsection{Results of experiments}

In this section, we present the experiment results for the first 12 time points ($\Theta=12$) in all scenarios. 

\paragraph{Experiment 1 - Feature identification accuracy} 

\begin{figure}[!htbp]
	 	
		\centerline{\includegraphics[height=21pc]{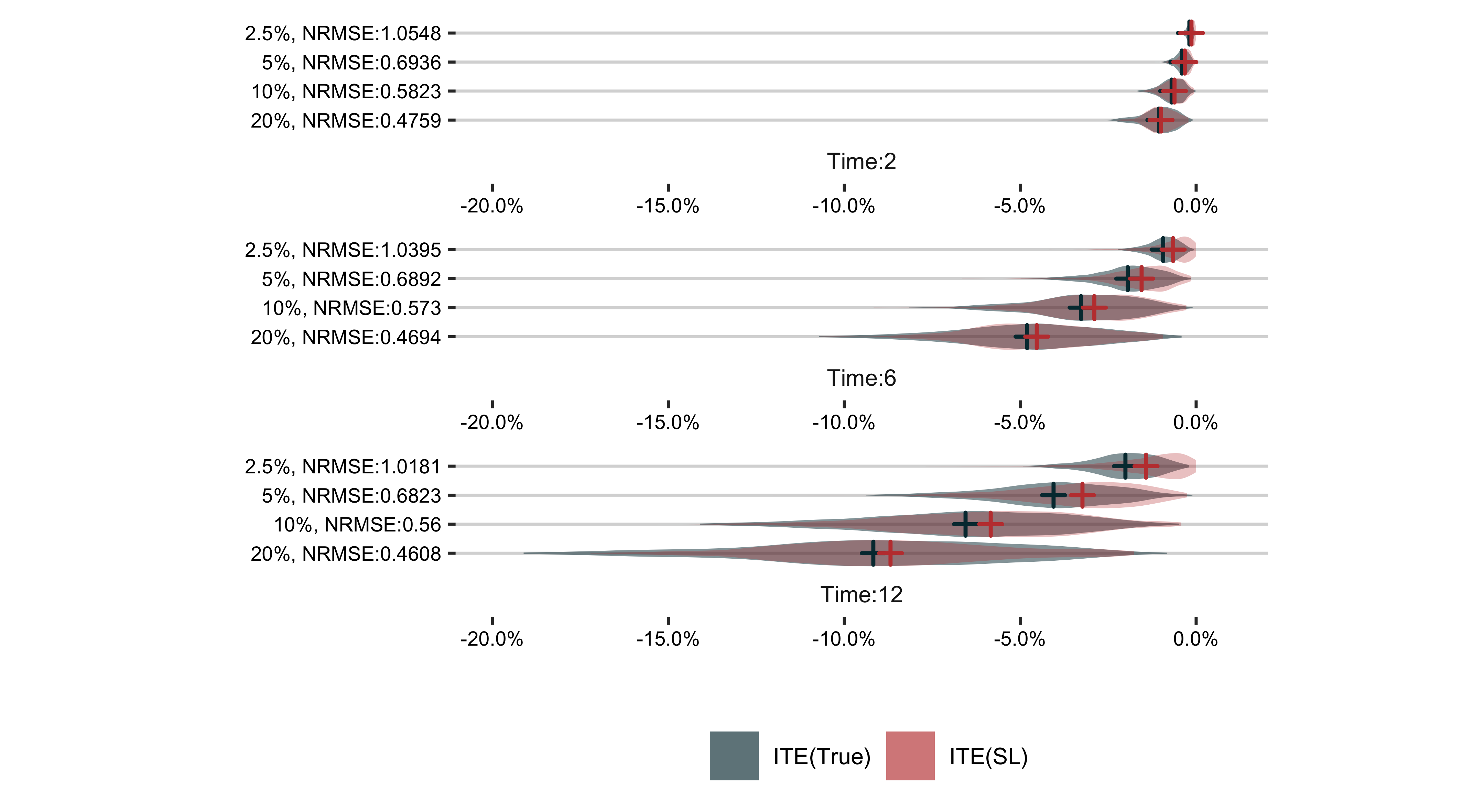}}\quad
		\caption{\fontsize{8pt}{10pt}\selectfont Distribution of true ITE (grey violin plots) and estimated effects from Super-learner (red violin plots). The vertical axis shows the  event rates and NRMSE of the estimation. The horizontal axis shows the value of ITE. The grey and red cross marks indicate the true and estimated average treatment effects.  We depicted the ITEs at time 1,6 and 12 from one of 50 simulations under default settings of sample size($N=3,000$), low confounding level($\beta=0.5$) and feature dimension($D=10$). 
		Abbreviation: ITE: Individual Treatment Effect, NRMSE: Normalised Root Mean Square Error}
		\label{ITEdist}
	\end{figure}

	We start by examining the distribution of the initial ITE estimations under different effect sizes. Figure \ref{ITEdist} depicts the true and estimated ITEs using samples under default scenarios with $\mathcal{R}=20\%, 10\%, 5\%$ and $2.5\%$ \footnote{We did not depict the ITE distributions under other scenarios (as they all have the same event rate by design) or the distributions after TMLE (as TMLE is not used for correcting individual estimations). In Appendix B, we can see the NRMSE after TMLE is worse than the NRMSE directly from the Super-learner estimation.}. Overall, we found  ITEs disperse over time with increasing effect heterogeneity. The scale of the effect heterogeneity is significantly higher in high event rate scenarios. When $t=12$, the ATE (indicated by cross marks) for $\mathcal{R}=20\%$ is 8.4 times higher than the $\mathcal{R}=2.5\%$ ($-11.8\%$ and $-1.4\%$).

	The accuracy of initial ITE estimator (in terms of NRMSE) is examined across different values of $N, D, \beta$ and $\mathcal{R}$. We notice that higher levels of confounding ($\beta$) raise NRMSE only by a small amount, for example: the NRMSE($t=12$) is 0.482 at $\beta=0$ and 0.521 at $\beta=2$, but the $\%$bias($t=12$) of the ATE is $8.41\%$ and $24.55\%$ respectively.

	Higher dimensions (D) also result in worse NRMSE, for example: NRMSE($t=12$) doubles when $D=30$ compared to $D=8$ given default values for other variables. However, the negative impact of higher dimensions can be offset by increasing the sample size. We observe NRMSE was halved as $N$ increased from $1,000$ to $15,000$. Lastly, lower event rates were related to higher NRMSE, which doubled from  $\mathcal{R}=20\%$ to $2.5\%$. It is also worth noting that the aforementioned trend also holds for the $\%$bias of ATE (Indicated by the cross marks in Figure \ref{ITEdist}). 
	
	The accuracy of feature identification is most affected by data dimension (D), and sample size (N). Given default values of event rate and confounding, Table \ref{tab1} illustrates a larger sample size is required to correctly identify all contributing features. When the sample size is $3,000$, BART is the best performing algorithm in terms of PPV and TPR. For $D>10$, the PPV and TPR drop below $50\%$ across all algorithms. At the largest sample size($N=15,000$), the performance of BART was lifted significantly after $D>10$. 

\begin{table*}[!htbp]%
	\caption{Diagnostic statistics for contributing feature identification ($\%$) \label{tab1}}
	\centering
	\fontsize{8pt}{10pt}\selectfont
	\begin{tabular}{cccccccccc|ccccccc}
	\toprule
	\multicolumn{5}{c}{} & \multicolumn{5}{c|}{\textbf{TPR Decomposition}} & \multicolumn{7}{c}{\textbf{$\Delta$: $N=9,000$ to $N=15,000$}} \\
	\cmidrule{6-17}\textbf{$N$} &   & \textbf{$D$} & \textbf{PPV} & \textbf{TPR} & \textbf{$X_1$} & \textbf{$X_2$} & \textbf{$X_3$} & \textbf{$X_4$} & \textbf{$X_5$} & \textbf{PPV} & \textbf{TPR} & \textbf{$X_1$} & \textbf{$X_2$} & \textbf{$X_3$} & \textbf{$X_4$} & \textbf{$X_5$} \\
	\midrule
	\multirow{20}[2]{*}{\begin{sideways}\textbf{15000}\end{sideways}} & \multicolumn{1}{c}{\multirow{5}[1]{*}{\begin{sideways}\textbf{Adaptive Lasso}\end{sideways}}} & 8 & \cellcolor[rgb]{ .949,  .949,  .949}\textcolor[rgb]{ .329,  .51,  .208}{\textbf{89}} & \cellcolor[rgb]{ .949,  .949,  .949}55 & 25 & 25 & 50 & 75 & \textbf{100} & \cellcolor[rgb]{ .949,  .949,  .949}26 & \cellcolor[rgb]{ .949,  .949,  .949}8 & \textcolor[rgb]{ .612,  0,  .024}{-25} & \textcolor[rgb]{ .612,  0,  .024}{-30} & \textcolor[rgb]{ .753,  0,  0}{25} & 25 & 45 \\
	  &   & 10 & \cellcolor[rgb]{ .949,  .949,  .949}68 & \cellcolor[rgb]{ .949,  .949,  .949}\textcolor[rgb]{ .329,  .51,  .208}{\textbf{50}} & 50 & \textbf{75} & 50 & 25 & 50 & \cellcolor[rgb]{ .949,  .949,  .949}14 & \cellcolor[rgb]{ .949,  .949,  .949}2 & \textcolor[rgb]{ .612,  0,  .024}{-10} & 20 & 45 & \textcolor[rgb]{ .612,  0,  .024}{-25} & \textcolor[rgb]{ .612,  0,  .024}{-20} \\
	  &   & 14 & \cellcolor[rgb]{ .949,  .949,  .949}42 & \cellcolor[rgb]{ .949,  .949,  .949}35 & 50 & \textbf{75} & 0 & 0 & 50 & \cellcolor[rgb]{ .949,  .949,  .949}\textcolor[rgb]{ .612,  0,  .024}{-3} & \cellcolor[rgb]{ .949,  .949,  .949}\textcolor[rgb]{ .612,  0,  .024}{-5} & 15 & 20 & \textcolor[rgb]{ .612,  0,  .024}{-30} & \textcolor[rgb]{ .612,  0,  .024}{-40} & 10 \\
	  &   & 20 & \cellcolor[rgb]{ .949,  .949,  .949}28 & \cellcolor[rgb]{ .949,  .949,  .949}30 & 0 & \textbf{75} & 25 & 25 & 25 & \cellcolor[rgb]{ .949,  .949,  .949}\textcolor[rgb]{ .612,  0,  .024}{-2} & \cellcolor[rgb]{ .949,  .949,  .949}0 & \textcolor[rgb]{ .612,  0,  .024}{-25} & 35 & 10 & \textcolor[rgb]{ .612,  0,  .024}{-15} & \textcolor[rgb]{ .612,  0,  .024}{-5} \\
	  &   & 30 & \cellcolor[rgb]{ .949,  .949,  .949}13 & \cellcolor[rgb]{ .949,  .949,  .949}20 & \textbf{50} & 0 & 25 & 25 & 0 & \cellcolor[rgb]{ .949,  .949,  .949}\textcolor[rgb]{ .612,  0,  .024}{-10} & \cellcolor[rgb]{ .949,  .949,  .949}\textcolor[rgb]{ .612,  0,  .024}{-4} & 25 & \textcolor[rgb]{ .612,  0,  .024}{-15} & 15 & 0 & \textcolor[rgb]{ .612,  0,  .024}{-45} \\
	  & \multirow{5}[0]{*}{\begin{sideways}\textbf{BART}\end{sideways}} & \cellcolor[rgb]{ .851,  .851,  .851}8 & \cellcolor[rgb]{ .851,  .851,  .851}83 & \cellcolor[rgb]{ .851,  .851,  .851}55 & \cellcolor[rgb]{ .851,  .851,  .851}50 & \cellcolor[rgb]{ .851,  .851,  .851}\textbf{100} & \cellcolor[rgb]{ .851,  .851,  .851}25 & \cellcolor[rgb]{ .851,  .851,  .851}25 & \cellcolor[rgb]{ .851,  .851,  .851}\textbf{100} & \cellcolor[rgb]{ .851,  .851,  .851}6 & \cellcolor[rgb]{ .851,  .851,  .851}3 & \cellcolor[rgb]{ .851,  .851,  .851}30 & \cellcolor[rgb]{ .851,  .851,  .851}5 & \cellcolor[rgb]{ .851,  .851,  .851}\textcolor[rgb]{ .612,  0,  .024}{-5} & \cellcolor[rgb]{ .851,  .851,  .851}\textcolor[rgb]{ .612,  0,  .024}{-15} & \cellcolor[rgb]{ .851,  .851,  .851}0 \\
	  &   & \cellcolor[rgb]{ .851,  .851,  .851}10 & \cellcolor[rgb]{ .851,  .851,  .851}\textcolor[rgb]{ .329,  .51,  .208}{\textbf{86}} & \cellcolor[rgb]{ .851,  .851,  .851}\textcolor[rgb]{ .329,  .51,  .208}{\textbf{50}} & \cellcolor[rgb]{ .851,  .851,  .851}25 & \cellcolor[rgb]{ .851,  .851,  .851}\textbf{100} & \cellcolor[rgb]{ .851,  .851,  .851}25 & \cellcolor[rgb]{ .851,  .851,  .851}25 & \cellcolor[rgb]{ .851,  .851,  .851}\textbf{100} & \cellcolor[rgb]{ .851,  .851,  .851}11 & \cellcolor[rgb]{ .851,  .851,  .851}\textcolor[rgb]{ .612,  0,  .024}{-11} & \cellcolor[rgb]{ .851,  .851,  .851}\textcolor[rgb]{ .612,  0,  .024}{-20} & \cellcolor[rgb]{ .851,  .851,  .851}5 & \cellcolor[rgb]{ .851,  .851,  .851}\textcolor[rgb]{ .612,  0,  .024}{-5} & \cellcolor[rgb]{ .851,  .851,  .851}\textcolor[rgb]{ .612,  0,  .024}{-35} & \cellcolor[rgb]{ .851,  .851,  .851}0 \\
	  &   & \cellcolor[rgb]{ .851,  .851,  .851}14 & \cellcolor[rgb]{ .851,  .851,  .851}\textcolor[rgb]{ .329,  .51,  .208}{\textbf{52}} & \cellcolor[rgb]{ .851,  .851,  .851}\textcolor[rgb]{ .329,  .51,  .208}{\textbf{75}} & \cellcolor[rgb]{ .851,  .851,  .851}\textbf{100} & \cellcolor[rgb]{ .851,  .851,  .851}\textbf{100} & \cellcolor[rgb]{ .851,  .851,  .851}50 & \cellcolor[rgb]{ .851,  .851,  .851}25 & \cellcolor[rgb]{ .851,  .851,  .851}\textbf{100} & \cellcolor[rgb]{ .851,  .851,  .851}\textcolor[rgb]{ .612,  0,  .024}{-10} & \cellcolor[rgb]{ .851,  .851,  .851}30 & \cellcolor[rgb]{ .851,  .851,  .851}75 & \cellcolor[rgb]{ .851,  .851,  .851}15 & \cellcolor[rgb]{ .851,  .851,  .851}40 & \cellcolor[rgb]{ .851,  .851,  .851}10 & \cellcolor[rgb]{ .851,  .851,  .851}10 \\
	  &   & \cellcolor[rgb]{ .851,  .851,  .851}20 & \cellcolor[rgb]{ .851,  .851,  .851}\textcolor[rgb]{ .329,  .51,  .208}{\textbf{43}} & \cellcolor[rgb]{ .851,  .851,  .851}\textcolor[rgb]{ .329,  .51,  .208}{\textbf{65}} & \cellcolor[rgb]{ .851,  .851,  .851}25 & \cellcolor[rgb]{ .851,  .851,  .851}\textbf{100} & \cellcolor[rgb]{ .851,  .851,  .851}50 & \cellcolor[rgb]{ .851,  .851,  .851}50 & \cellcolor[rgb]{ .851,  .851,  .851}\textbf{100} & \cellcolor[rgb]{ .851,  .851,  .851}1 & \cellcolor[rgb]{ .851,  .851,  .851}25 & \cellcolor[rgb]{ .851,  .851,  .851}5 & \cellcolor[rgb]{ .851,  .851,  .851}25 & \cellcolor[rgb]{ .851,  .851,  .851}30 & \cellcolor[rgb]{ .851,  .851,  .851}30 & \cellcolor[rgb]{ .851,  .851,  .851}35 \\
	  &   & \cellcolor[rgb]{ .851,  .851,  .851}30 & \cellcolor[rgb]{ .851,  .851,  .851}\textcolor[rgb]{ .329,  .51,  .208}{\textbf{25}} & \cellcolor[rgb]{ .851,  .851,  .851}\textcolor[rgb]{ .329,  .51,  .208}{\textbf{55}} & \cellcolor[rgb]{ .851,  .851,  .851}25 & \cellcolor[rgb]{ .851,  .851,  .851}\textbf{100} & \cellcolor[rgb]{ .851,  .851,  .851}25 & \cellcolor[rgb]{ .851,  .851,  .851}50 & \cellcolor[rgb]{ .851,  .851,  .851}\textbf{100} & \cellcolor[rgb]{ .851,  .851,  .851}1 & \cellcolor[rgb]{ .851,  .851,  .851}31 & \cellcolor[rgb]{ .851,  .851,  .851}10 & \cellcolor[rgb]{ .851,  .851,  .851}60 & \cellcolor[rgb]{ .851,  .851,  .851}0 & \cellcolor[rgb]{ .851,  .851,  .851}30 & \cellcolor[rgb]{ .851,  .851,  .851}55 \\
	  & \multirow{5}[0]{*}{\begin{sideways}\textbf{Causal Forest}\end{sideways}} & 8 & \cellcolor[rgb]{ .949,  .949,  .949}57 & \cellcolor[rgb]{ .949,  .949,  .949}\textcolor[rgb]{ .329,  .51,  .208}{\textbf{60}} & \textbf{75} & 50 & \textbf{75} & \textbf{75} & 25 & \cellcolor[rgb]{ .949,  .949,  .949}\textcolor[rgb]{ .612,  0,  .024}{-1} & \cellcolor[rgb]{ .949,  .949,  .949}1 & 0 & 5 & 5 & 10 & \textcolor[rgb]{ .612,  0,  .024}{-15} \\
	  &   & 10 & \cellcolor[rgb]{ .949,  .949,  .949}39 & \cellcolor[rgb]{ .949,  .949,  .949}40 & \textbf{75} & 0 & 50 & 50 & 25 & \cellcolor[rgb]{ .949,  .949,  .949}\textcolor[rgb]{ .612,  0,  .024}{-5} & \cellcolor[rgb]{ .949,  .949,  .949}\textcolor[rgb]{ .612,  0,  .024}{-16} & 15 & \textcolor[rgb]{ .612,  0,  .024}{-45} & \textcolor[rgb]{ .612,  0,  .024}{-15} & \textcolor[rgb]{ .612,  0,  .024}{-15} & \textcolor[rgb]{ .612,  0,  .024}{-20} \\
	  &   & 14 & \cellcolor[rgb]{ .949,  .949,  .949}48 & \cellcolor[rgb]{ .949,  .949,  .949}45 & 50 & 25 & 50 & \textbf{75} & 25 & \cellcolor[rgb]{ .949,  .949,  .949}13 & \cellcolor[rgb]{ .949,  .949,  .949}0 & 0 & \textcolor[rgb]{ .612,  0,  .024}{-5} & \textcolor[rgb]{ .612,  0,  .024}{-5} & 25 & \textcolor[rgb]{ .612,  0,  .024}{-15} \\
	  &   & 20 & \cellcolor[rgb]{ .949,  .949,  .949}33 & \cellcolor[rgb]{ .949,  .949,  .949}15 & 0 & 25 & 0 & \textbf{50} & 0 & \cellcolor[rgb]{ .949,  .949,  .949}7 & \cellcolor[rgb]{ .949,  .949,  .949}\textcolor[rgb]{ .612,  0,  .024}{-18} & \textcolor[rgb]{ .612,  0,  .024}{-35} & 10 & \textcolor[rgb]{ .612,  0,  .024}{-45} & 20 & \textcolor[rgb]{ .612,  0,  .024}{-40} \\
	  &   & 30 & \cellcolor[rgb]{ .949,  .949,  .949}17 & \cellcolor[rgb]{ .949,  .949,  .949}10 & 0 & \textbf{25} & 0 & \textbf{25} & 0 & \cellcolor[rgb]{ .949,  .949,  .949}4 & \cellcolor[rgb]{ .949,  .949,  .949}\textcolor[rgb]{ .612,  0,  .024}{-7} & \textcolor[rgb]{ .612,  0,  .024}{-5} & 0 & \textcolor[rgb]{ .612,  0,  .024}{-15} & 0 & \textcolor[rgb]{ .612,  0,  .024}{-15} \\
	  & \multirow{5}[1]{*}{\begin{sideways}\textbf{Elastic Net}\end{sideways}} & \cellcolor[rgb]{ .851,  .851,  .851}8 & \cellcolor[rgb]{ .851,  .851,  .851}\textcolor[rgb]{ .329,  .51,  .208}{\textbf{89}} & \cellcolor[rgb]{ .851,  .851,  .851}55 & \cellcolor[rgb]{ .851,  .851,  .851}25 & \cellcolor[rgb]{ .851,  .851,  .851}25 & \cellcolor[rgb]{ .851,  .851,  .851}50 & \cellcolor[rgb]{ .851,  .851,  .851}75 & \cellcolor[rgb]{ .851,  .851,  .851}\textbf{100} & \cellcolor[rgb]{ .851,  .851,  .851}25 & \cellcolor[rgb]{ .851,  .851,  .851}11 & \cellcolor[rgb]{ .851,  .851,  .851}\textcolor[rgb]{ .612,  0,  .024}{-25} & \cellcolor[rgb]{ .851,  .851,  .851}\textcolor[rgb]{ .612,  0,  .024}{-30} & \cellcolor[rgb]{ .851,  .851,  .851}25 & \cellcolor[rgb]{ .851,  .851,  .851}30 & \cellcolor[rgb]{ .851,  .851,  .851}55 \\
	  &   & \cellcolor[rgb]{ .851,  .851,  .851}10 & \cellcolor[rgb]{ .851,  .851,  .851}68 & \cellcolor[rgb]{ .851,  .851,  .851}50 & \cellcolor[rgb]{ .851,  .851,  .851}50 & \cellcolor[rgb]{ .851,  .851,  .851}\textbf{75} & \cellcolor[rgb]{ .851,  .851,  .851}50 & \cellcolor[rgb]{ .851,  .851,  .851}25 & \cellcolor[rgb]{ .851,  .851,  .851}50 & \cellcolor[rgb]{ .851,  .851,  .851}14 & \cellcolor[rgb]{ .851,  .851,  .851}0 & \cellcolor[rgb]{ .851,  .851,  .851}\textcolor[rgb]{ .612,  0,  .024}{-10} & \cellcolor[rgb]{ .851,  .851,  .851}15 & \cellcolor[rgb]{ .851,  .851,  .851}40 & \cellcolor[rgb]{ .851,  .851,  .851}\textcolor[rgb]{ .612,  0,  .024}{-25} & \cellcolor[rgb]{ .851,  .851,  .851}\textcolor[rgb]{ .612,  0,  .024}{-20} \\
	  &   & \cellcolor[rgb]{ .851,  .851,  .851}14 & \cellcolor[rgb]{ .851,  .851,  .851}42 & \cellcolor[rgb]{ .851,  .851,  .851}35 & \cellcolor[rgb]{ .851,  .851,  .851}50 & \cellcolor[rgb]{ .851,  .851,  .851}\textbf{75} & \cellcolor[rgb]{ .851,  .851,  .851}0 & \cellcolor[rgb]{ .851,  .851,  .851}0 & \cellcolor[rgb]{ .851,  .851,  .851}50 & \cellcolor[rgb]{ .851,  .851,  .851}\textcolor[rgb]{ .612,  0,  .024}{-3} & \cellcolor[rgb]{ .851,  .851,  .851}\textcolor[rgb]{ .612,  0,  .024}{-5} & \cellcolor[rgb]{ .851,  .851,  .851}15 & \cellcolor[rgb]{ .851,  .851,  .851}20 & \cellcolor[rgb]{ .851,  .851,  .851}\textcolor[rgb]{ .612,  0,  .024}{-30} & \cellcolor[rgb]{ .851,  .851,  .851}\textcolor[rgb]{ .612,  0,  .024}{-40} & \cellcolor[rgb]{ .851,  .851,  .851}10 \\
	  &   & \cellcolor[rgb]{ .851,  .851,  .851}20 & \cellcolor[rgb]{ .851,  .851,  .851}28 & \cellcolor[rgb]{ .851,  .851,  .851}30 & \cellcolor[rgb]{ .851,  .851,  .851}0 & \cellcolor[rgb]{ .851,  .851,  .851}\textbf{75} & \cellcolor[rgb]{ .851,  .851,  .851}25 & \cellcolor[rgb]{ .851,  .851,  .851}25 & \cellcolor[rgb]{ .851,  .851,  .851}25 & \cellcolor[rgb]{ .851,  .851,  .851}\textcolor[rgb]{ .612,  0,  .024}{-2} & \cellcolor[rgb]{ .851,  .851,  .851}0 & \cellcolor[rgb]{ .851,  .851,  .851}\textcolor[rgb]{ .612,  0,  .024}{-25} & \cellcolor[rgb]{ .851,  .851,  .851}35 & \cellcolor[rgb]{ .851,  .851,  .851}10 & \cellcolor[rgb]{ .851,  .851,  .851}\textcolor[rgb]{ .612,  0,  .024}{-15} & \cellcolor[rgb]{ .851,  .851,  .851}\textcolor[rgb]{ .612,  0,  .024}{-5} \\
	  &   & \cellcolor[rgb]{ .851,  .851,  .851}30 & \cellcolor[rgb]{ .851,  .851,  .851}11 & \cellcolor[rgb]{ .851,  .851,  .851}20 & \cellcolor[rgb]{ .851,  .851,  .851}\textbf{50} & \cellcolor[rgb]{ .851,  .851,  .851}0 & \cellcolor[rgb]{ .851,  .851,  .851}25 & \cellcolor[rgb]{ .851,  .851,  .851}25 & \cellcolor[rgb]{ .851,  .851,  .851}0 & \cellcolor[rgb]{ .851,  .851,  .851}\textcolor[rgb]{ .612,  0,  .024}{-12} & \cellcolor[rgb]{ .851,  .851,  .851}\textcolor[rgb]{ .612,  0,  .024}{-3} & \cellcolor[rgb]{ .851,  .851,  .851}25 & \cellcolor[rgb]{ .851,  .851,  .851}\textcolor[rgb]{ .612,  0,  .024}{-15} & \cellcolor[rgb]{ .851,  .851,  .851}15 & \cellcolor[rgb]{ .851,  .851,  .851}0 & \cellcolor[rgb]{ .851,  .851,  .851}\textcolor[rgb]{ .612,  0,  .024}{-40} \\
	\midrule
	  &   &   &   &   &   &   &   &   &   & \multicolumn{7}{c}{\textbf{$\Delta$: $N=3,000$ to $N=9,000$}} \\
	\midrule
	\multirow{20}[2]{*}{\begin{sideways}\textbf{3000}\end{sideways}} & \multicolumn{1}{c}{\multirow{5}[1]{*}{\begin{sideways}\textbf{Adaptive Lasso}\end{sideways}}} & 8 & \cellcolor[rgb]{ .949,  .949,  .949}62 & \cellcolor[rgb]{ .949,  .949,  .949}47 & \textbf{57} & 50 & 27 & 50 & 50 & \cellcolor[rgb]{ .949,  .949,  .949}2 & \cellcolor[rgb]{ .949,  .949,  .949}0 & \textcolor[rgb]{ .612,  0,  .024}{-7} & 5 & \textcolor[rgb]{ .612,  0,  .024}{-2} & 0 & 5 \\
	  &   & 10 & \cellcolor[rgb]{ .949,  .949,  .949}50 & \cellcolor[rgb]{ .949,  .949,  .949}47 & 30 & 57 & 37 & 40 & \textbf{70} & \cellcolor[rgb]{ .949,  .949,  .949}4 & \cellcolor[rgb]{ .949,  .949,  .949}1 & 30 & \textcolor[rgb]{ .612,  0,  .024}{-2} & \textcolor[rgb]{ .612,  0,  .024}{-32} & 10 & 0 \\
	  &   & 14 & \cellcolor[rgb]{ .949,  .949,  .949}43 & \cellcolor[rgb]{ .949,  .949,  .949}31 & 23 & \textbf{43} & 33 & 27 & 27 & \cellcolor[rgb]{ .949,  .949,  .949}2 & \cellcolor[rgb]{ .949,  .949,  .949}9 & 12 & 12 & \textcolor[rgb]{ .612,  0,  .024}{-3} & 13 & 13 \\
	  &   & 20 & \cellcolor[rgb]{ .949,  .949,  .949}26 & \cellcolor[rgb]{ .949,  .949,  .949}21 & \textbf{27} & 10 & 23 & 20 & 23 & \cellcolor[rgb]{ .949,  .949,  .949}3 & \cellcolor[rgb]{ .949,  .949,  .949}9 & \textcolor[rgb]{ .612,  0,  .024}{-2} & 30 & \textcolor[rgb]{ .612,  0,  .024}{-8} & 20 & 7 \\
	  &   & 30 & \cellcolor[rgb]{ .949,  .949,  .949}15 & \cellcolor[rgb]{ .949,  .949,  .949}12 & 10 & \textbf{13} & 10 & \textbf{13} & \textbf{13} & \cellcolor[rgb]{ .949,  .949,  .949}8 & \cellcolor[rgb]{ .949,  .949,  .949}12 & 15 & 2 & 0 & 12 & 32 \\
	  & \multirow{5}[0]{*}{\begin{sideways}\textbf{BART}\end{sideways}} & \cellcolor[rgb]{ .851,  .851,  .851}8 & \cellcolor[rgb]{ .851,  .851,  .851}\textcolor[rgb]{ .329,  .51,  .208}{\textbf{82}} & \cellcolor[rgb]{ .851,  .851,  .851}\textcolor[rgb]{ .329,  .51,  .208}{\textbf{54}} & \cellcolor[rgb]{ .851,  .851,  .851}37 & \cellcolor[rgb]{ .851,  .851,  .851}\textbf{93} & \cellcolor[rgb]{ .851,  .851,  .851}3 & \cellcolor[rgb]{ .851,  .851,  .851}43 & \cellcolor[rgb]{ .851,  .851,  .851}\textbf{93} & \cellcolor[rgb]{ .851,  .851,  .851}\textcolor[rgb]{ .612,  0,  .024}{-5} & \cellcolor[rgb]{ .851,  .851,  .851}\textcolor[rgb]{ .612,  0,  .024}{-2} & \cellcolor[rgb]{ .851,  .851,  .851}\textcolor[rgb]{ .612,  0,  .024}{-17} & \cellcolor[rgb]{ .851,  .851,  .851}2 & \cellcolor[rgb]{ .851,  .851,  .851}2 & \cellcolor[rgb]{ .851,  .851,  .851}\textcolor[rgb]{ .612,  0,  .024}{-3} & \cellcolor[rgb]{ .851,  .851,  .851}7 \\
	  &   & \cellcolor[rgb]{ .851,  .851,  .851}10 & \cellcolor[rgb]{ .851,  .851,  .851}\textcolor[rgb]{ .329,  .51,  .208}{\textbf{66}} & \cellcolor[rgb]{ .851,  .851,  .851}\textcolor[rgb]{ .329,  .51,  .208}{\textbf{47}} & \cellcolor[rgb]{ .851,  .851,  .851}40 & \cellcolor[rgb]{ .851,  .851,  .851}\textbf{83} & \cellcolor[rgb]{ .851,  .851,  .851}10 & \cellcolor[rgb]{ .851,  .851,  .851}40 & \cellcolor[rgb]{ .851,  .851,  .851}63 & \cellcolor[rgb]{ .851,  .851,  .851}9 & \cellcolor[rgb]{ .851,  .851,  .851}13 & \cellcolor[rgb]{ .851,  .851,  .851}5 & \cellcolor[rgb]{ .851,  .851,  .851}12 & \cellcolor[rgb]{ .851,  .851,  .851}\textcolor[rgb]{ .612,  0,  .024}{-5} & \cellcolor[rgb]{ .851,  .851,  .851}15 & \cellcolor[rgb]{ .851,  .851,  .851}37 \\
	  &   & \cellcolor[rgb]{ .851,  .851,  .851}14 & \cellcolor[rgb]{ .851,  .851,  .851}\textcolor[rgb]{ .329,  .51,  .208}{\textbf{50}} & \cellcolor[rgb]{ .851,  .851,  .851}\textcolor[rgb]{ .329,  .51,  .208}{\textbf{43}} & \cellcolor[rgb]{ .851,  .851,  .851}40 & \cellcolor[rgb]{ .851,  .851,  .851}\textbf{57} & \cellcolor[rgb]{ .851,  .851,  .851}20 & \cellcolor[rgb]{ .851,  .851,  .851}47 & \cellcolor[rgb]{ .851,  .851,  .851}53 & \cellcolor[rgb]{ .851,  .851,  .851}13 & \cellcolor[rgb]{ .851,  .851,  .851}2 & \cellcolor[rgb]{ .851,  .851,  .851}\textcolor[rgb]{ .612,  0,  .024}{-15} & \cellcolor[rgb]{ .851,  .851,  .851}28 & \cellcolor[rgb]{ .851,  .851,  .851}\textcolor[rgb]{ .612,  0,  .024}{-10} & \cellcolor[rgb]{ .851,  .851,  .851}\textcolor[rgb]{ .612,  0,  .024}{-32} & \cellcolor[rgb]{ .851,  .851,  .851}37 \\
	  &   & \cellcolor[rgb]{ .851,  .851,  .851}20 & \cellcolor[rgb]{ .851,  .851,  .851}\textcolor[rgb]{ .329,  .51,  .208}{\textbf{32}} & \cellcolor[rgb]{ .851,  .851,  .851}\textcolor[rgb]{ .329,  .51,  .208}{\textbf{33}} & \cellcolor[rgb]{ .851,  .851,  .851}30 & \cellcolor[rgb]{ .851,  .851,  .851}\textbf{47} & \cellcolor[rgb]{ .851,  .851,  .851}13 & \cellcolor[rgb]{ .851,  .851,  .851}43 & \cellcolor[rgb]{ .851,  .851,  .851}30 & \cellcolor[rgb]{ .851,  .851,  .851}10 & \cellcolor[rgb]{ .851,  .851,  .851}7 & \cellcolor[rgb]{ .851,  .851,  .851}\textcolor[rgb]{ .612,  0,  .024}{-10} & \cellcolor[rgb]{ .851,  .851,  .851}28 & \cellcolor[rgb]{ .851,  .851,  .851}7 & \cellcolor[rgb]{ .851,  .851,  .851}\textcolor[rgb]{ .612,  0,  .024}{-23} & \cellcolor[rgb]{ .851,  .851,  .851}35 \\
	  &   & \cellcolor[rgb]{ .851,  .851,  .851}30 & \cellcolor[rgb]{ .851,  .851,  .851}\textcolor[rgb]{ .329,  .51,  .208}{\textbf{21}} & \cellcolor[rgb]{ .851,  .851,  .851}\textcolor[rgb]{ .329,  .51,  .208}{\textbf{24}} & \cellcolor[rgb]{ .851,  .851,  .851}27 & \cellcolor[rgb]{ .851,  .851,  .851}\textbf{37} & \cellcolor[rgb]{ .851,  .851,  .851}13 & \cellcolor[rgb]{ .851,  .851,  .851}13 & \cellcolor[rgb]{ .851,  .851,  .851}30 & \cellcolor[rgb]{ .851,  .851,  .851}4 & \cellcolor[rgb]{ .851,  .851,  .851}0 & \cellcolor[rgb]{ .851,  .851,  .851}\textcolor[rgb]{ .612,  0,  .024}{-12} & \cellcolor[rgb]{ .851,  .851,  .851}3 & \cellcolor[rgb]{ .851,  .851,  .851}\textcolor[rgb]{ .612,  0,  .024}{-13} & \cellcolor[rgb]{ .851,  .851,  .851}7 & \cellcolor[rgb]{ .851,  .851,  .851}15 \\
	  & \multirow{5}[0]{*}{\begin{sideways}\textbf{Causal Forest}\end{sideways}} & 8 & \cellcolor[rgb]{ .949,  .949,  .949}65 & \cellcolor[rgb]{ .949,  .949,  .949}45 & 47 & 43 & \textbf{50} & 43 & 40 & \cellcolor[rgb]{ .949,  .949,  .949}\textcolor[rgb]{ .612,  0,  .024}{-7} & \cellcolor[rgb]{ .949,  .949,  .949}14 & 28 & 2 & 20 & 22 & 0 \\
	  &   & 10 & \cellcolor[rgb]{ .949,  .949,  .949}54 & \cellcolor[rgb]{ .949,  .949,  .949}43 & \textbf{47} & 30 & 53 & \textbf{47} & 37 & \cellcolor[rgb]{ .949,  .949,  .949}\textcolor[rgb]{ .612,  0,  .024}{-11} & \cellcolor[rgb]{ .949,  .949,  .949}13 & 13 & 15 & 12 & 18 & 8 \\
	  &   & 14 & \cellcolor[rgb]{ .949,  .949,  .949}38 & \cellcolor[rgb]{ .949,  .949,  .949}39 & 37 & 47 & \textbf{53} & 23 & 37 & \cellcolor[rgb]{ .949,  .949,  .949}\textcolor[rgb]{ .612,  0,  .024}{-3} & \cellcolor[rgb]{ .949,  .949,  .949}6 & 13 & \textcolor[rgb]{ .612,  0,  .024}{-17} & 2 & 27 & 3 \\
	  &   & 20 & \cellcolor[rgb]{ .949,  .949,  .949}23 & \cellcolor[rgb]{ .949,  .949,  .949}25 & 27 & 33 & 13 & 30 & 23 & \cellcolor[rgb]{ .949,  .949,  .949}3 & \cellcolor[rgb]{ .949,  .949,  .949}8 & 8 & \textcolor[rgb]{ .612,  0,  .024}{-18} & 32 & 0 & 17 \\
	  &   & 30 & \cellcolor[rgb]{ .949,  .949,  .949}12 & \cellcolor[rgb]{ .949,  .949,  .949}15 & 23 & 17 & 13 & 3 & 20 & \cellcolor[rgb]{ .949,  .949,  .949}1 & \cellcolor[rgb]{ .949,  .949,  .949}2 & \textcolor[rgb]{ .612,  0,  .024}{-18} & 8 & 2 & 22 & \textcolor[rgb]{ .612,  0,  .024}{-5} \\
	  & \multirow{5}[1]{*}{\begin{sideways}\textbf{Elastic Net}\end{sideways}} & \cellcolor[rgb]{ .851,  .851,  .851}8 & \cellcolor[rgb]{ .851,  .851,  .851}61 & \cellcolor[rgb]{ .851,  .851,  .851}45 & \cellcolor[rgb]{ .851,  .851,  .851}\textbf{53} & \cellcolor[rgb]{ .851,  .851,  .851}50 & \cellcolor[rgb]{ .851,  .851,  .851}27 & \cellcolor[rgb]{ .851,  .851,  .851}47 & \cellcolor[rgb]{ .851,  .851,  .851}50 & \cellcolor[rgb]{ .851,  .851,  .851}4 & \cellcolor[rgb]{ .851,  .851,  .851}\textcolor[rgb]{ .612,  0,  .024}{-1} & \cellcolor[rgb]{ .851,  .851,  .851}\textcolor[rgb]{ .612,  0,  .024}{-3} & \cellcolor[rgb]{ .851,  .851,  .851}5 & \cellcolor[rgb]{ .851,  .851,  .851}\textcolor[rgb]{ .612,  0,  .024}{-2} & \cellcolor[rgb]{ .851,  .851,  .851}\textcolor[rgb]{ .612,  0,  .024}{-2} & \cellcolor[rgb]{ .851,  .851,  .851}\textcolor[rgb]{ .612,  0,  .024}{-5} \\
	  &   & \cellcolor[rgb]{ .851,  .851,  .851}10 & \cellcolor[rgb]{ .851,  .851,  .851}51 & \cellcolor[rgb]{ .851,  .851,  .851}\textcolor[rgb]{ .329,  .51,  .208}{\textbf{47}} & \cellcolor[rgb]{ .851,  .851,  .851}33 & \cellcolor[rgb]{ .851,  .851,  .851}57 & \cellcolor[rgb]{ .851,  .851,  .851}37 & \cellcolor[rgb]{ .851,  .851,  .851}40 & \cellcolor[rgb]{ .851,  .851,  .851}\textbf{70} & \cellcolor[rgb]{ .851,  .851,  .851}2 & \cellcolor[rgb]{ .851,  .851,  .851}3 & \cellcolor[rgb]{ .851,  .851,  .851}27 & \cellcolor[rgb]{ .851,  .851,  .851}3 & \cellcolor[rgb]{ .851,  .851,  .851}\textcolor[rgb]{ .612,  0,  .024}{-27} & \cellcolor[rgb]{ .851,  .851,  .851}10 & \cellcolor[rgb]{ .851,  .851,  .851}0 \\
	  &   & \cellcolor[rgb]{ .851,  .851,  .851}14 & \cellcolor[rgb]{ .851,  .851,  .851}43 & \cellcolor[rgb]{ .851,  .851,  .851}33 & \cellcolor[rgb]{ .851,  .851,  .851}27 & \cellcolor[rgb]{ .851,  .851,  .851}\textbf{43} & \cellcolor[rgb]{ .851,  .851,  .851}37 & \cellcolor[rgb]{ .851,  .851,  .851}30 & \cellcolor[rgb]{ .851,  .851,  .851}30 & \cellcolor[rgb]{ .851,  .851,  .851}1 & \cellcolor[rgb]{ .851,  .851,  .851}7 & \cellcolor[rgb]{ .851,  .851,  .851}8 & \cellcolor[rgb]{ .851,  .851,  .851}12 & \cellcolor[rgb]{ .851,  .851,  .851}\textcolor[rgb]{ .612,  0,  .024}{-7} & \cellcolor[rgb]{ .851,  .851,  .851}10 & \cellcolor[rgb]{ .851,  .851,  .851}10 \\
	  &   & \cellcolor[rgb]{ .851,  .851,  .851}20 & \cellcolor[rgb]{ .851,  .851,  .851}26 & \cellcolor[rgb]{ .851,  .851,  .851}19 & \cellcolor[rgb]{ .851,  .851,  .851}\textbf{30} & \cellcolor[rgb]{ .851,  .851,  .851}10 & \cellcolor[rgb]{ .851,  .851,  .851}13 & \cellcolor[rgb]{ .851,  .851,  .851}20 & \cellcolor[rgb]{ .851,  .851,  .851}20 & \cellcolor[rgb]{ .851,  .851,  .851}3 & \cellcolor[rgb]{ .851,  .851,  .851}11 & \cellcolor[rgb]{ .851,  .851,  .851}\textcolor[rgb]{ .612,  0,  .024}{-5} & \cellcolor[rgb]{ .851,  .851,  .851}30 & \cellcolor[rgb]{ .851,  .851,  .851}2 & \cellcolor[rgb]{ .851,  .851,  .851}20 & \cellcolor[rgb]{ .851,  .851,  .851}10 \\
	  &   & \cellcolor[rgb]{ .851,  .851,  .851}30 & \cellcolor[rgb]{ .851,  .851,  .851}18 & \cellcolor[rgb]{ .851,  .851,  .851}15 & \cellcolor[rgb]{ .851,  .851,  .851}\textbf{23} & \cellcolor[rgb]{ .851,  .851,  .851}17 & \cellcolor[rgb]{ .851,  .851,  .851}10 & \cellcolor[rgb]{ .851,  .851,  .851}13 & \cellcolor[rgb]{ .851,  .851,  .851}10 & \cellcolor[rgb]{ .851,  .851,  .851}5 & \cellcolor[rgb]{ .851,  .851,  .851}8 & \cellcolor[rgb]{ .851,  .851,  .851}2 & \cellcolor[rgb]{ .851,  .851,  .851}\textcolor[rgb]{ .612,  0,  .024}{-2} & \cellcolor[rgb]{ .851,  .851,  .851}0 & \cellcolor[rgb]{ .851,  .851,  .851}12 & \cellcolor[rgb]{ .851,  .851,  .851}30 \\
	\bottomrule
	\end{tabular}%

\begin{tablenotes}
\item Each cell on the LHS of the vertical line is the diagnostics metrics. For each dimension and sample size, we highlight values with the best PPV and TRP in green. The TPR decomposition shows the percentage of times when the true feature is identified across all samples (we bold the best value on each row). On the RHS, we give the difference ($\Delta$) in metrics by increasing the sample size from a smaller value to a bigger one. All metrics are averaged across 50 samples. 
\item Abbreviations: PPV, Positive Predictive Value; TPR,  True Positive Rate; $N$, Sample Size; $D$, Feature Dimension.
\end{tablenotes}
\end{table*}

\paragraph{Experiment 2 - CATE estimation accuracy} 

We illustrate CATE estimation accuracy by averaging the $\%$Bias of 50 subgroups of size 300 for $X_2 \in [0,0.1)$ under default simulation settings. Panels (a) to (d) in Figure X illustrate the $\%$Bias as a function of follow-up time for different levels of confounding (a), dimension (b), event rate (c) and subgroup size (d). There is no notable change in bias over time, although standard errors increase with time due to smaller sample sizes (see Appendix X). As expected,  low confounding, low dimension or high event rate scenarios enjoy higher estimation accuracy. The subgroup size, on the other hand, is not a major factor for CATE accuracy(if the initial ITEs are accurately estimated, then taking the average of these ITEs will be accurate for any subgroup sizes). Both event rate and data dimension have an important impact on CATE accuracy. 

Panels (e-g) examine the impact of event rate in combination with confounding (e), dimension (f) and sample size(h). When $\mathcal{R}\geq 20\%$, the bias is contained within $10\%$ for all examined situations except the case of  $D=30$. For $\mathcal{R}\leq 5\%$, the CATE estimation quality is most affected by D with bias increasing significantly to $20\%-30\%$.  Sample size seems to have a small impact after controlling dimension and confounding level. Only at $\mathcal{R}=2.5\%$,  we see a significant uplift in bias for $N\leq 3,000$. 

Panels (i-j) examine the impact of dimension in combination with sample size (i) and confounding (j). By controlling $\mathcal{R}=10\%$, increasing the sample size is not very effective to reduce bias for high dimension scenarios. However, the bias is lower when a high-dimension scenario occurs with low confounding. A complete list of the results related to these experiments can be found in Appendix B.

\section{Case study}

	\subsection{Data Source}
	We provide a case study to illustrate the application and utility of this framework. We conducted a retrospective observational cohort study that estimates the effect of oral anticoagulants for the prevention of stroke for patients with newly diagnosed non-valvular atrial fibrillation (AF). This is part of a larger project on investigating the comparative effectiveness of anticoagulants that makes use of the UK's Clinical Practice Research Datalink (CPRD)\cite{Herrett:2015fc} and includes the primary care database (GOLD) linked to the Hospital Episode Statistics (HES) Admitted Patient Care, the HES Outpatient Data, the Office for National Statistics (ONS) Mortality data and the Patient Level index of Multiple Deprivation. The cohort consisted of patients with a diagnosis of AF (either in CPRD or HES datasets) between 1st-Mar-2011 and 31-July-2017, 18 years or over at time of first diagnosis, meeting CPRD's research quality indicators, registered with CPRD for at least one year before their initial diagnosis, and who had their records linked to HES data. Patients with a history of AF or with a record for a valve condition, prosthesis or procedure previous to initial diagnosis were excluded from the study. Oral anticoagulant exposure was determined from prescription records in CPRD by estimating the duration of a prescription as the total prescribed dose divided by a daily dose. A treatment was considered as  being discontinued if the gap between a prescription and a subsequent prescription (or, in the case of Vitamin K Antagonists (VKAs), a subsequent prescription or INR test) exceeds a grace period of 30 days. 

	We measured three types of outcomes: (1) death, (2) ischaemic event (which we simply refer to as stroke) and (3) major bleeding. Patients with any outcome under consideration has taken place within 15 days from the start date were removed from the study since these endpoints cannot be attributed to treatment effects. A patient is considered loss to follow-up if (1) dies; (2) leaves the practice or the practice leaves CPRD; (3) discontinues treatment; or (4) at the end of the study period; whichever happens first. The resulting sample contained 28,972 patients with 48 baseline covariates including age, gender, key comorbidities and medications and 3 indices of risk of the outcome at baseline, namely the Charlson index, $\mathrm{CHA}_2\mathrm{DS}_2$-$\mathrm{VASc}$ score and HAS-BLED Score. A table with the distribution of baseline covariates can be found in Appendix C.

	\subsection{Methodology}  As the true contributing feature is unknown, we assess which feature is most frequently identified across multiple random samples drawn from the whole dataset. In this study, we draw 30 bootstrap random samples of size $8,000$ from $28,972$ observations. An important covariate is expected to be identified in all 30 samples. 
	For visualisation of confidence interval, we calculate the one-side simultaneous confidence interval, $\mathrm{CI}^*$, as:
	\begin{eqnarray*}
		\mathrm{CI}^* = \mathrm{CI}_{\mathrm{upper}}-\hat\psi, 
	\end{eqnarray*}
	where $\mathrm{CI}_{\mathrm{upper}}$ is the upper $95\%$ simultaneous confidence interval and $\hat\psi$ is the estimated effect. 

	\subsection{Results of the case study}

	\begin{figure}[!htbp]
	\centerline{\includegraphics[height=20pc]{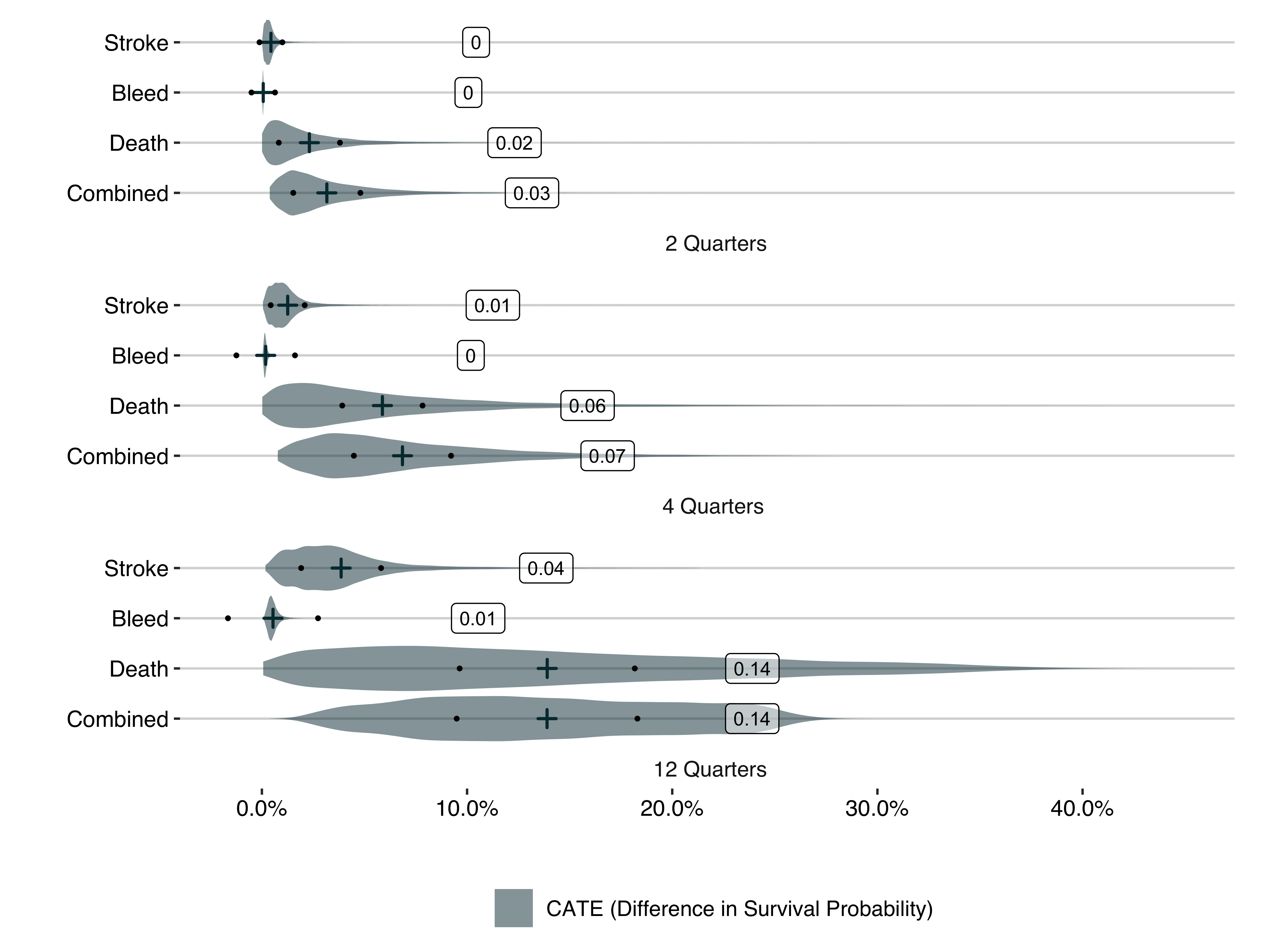}}\quad
	\caption{\fontsize{8pt}{10pt}\selectfont Distribution of individual treatment effect (grey violin plots) and average treatment effect estimations (grey cross marks and corresponding values in the squares) with $95\%$ simultaneous confidence interval (black dots).}
	\label{ITEdist2}
	\end{figure} 

	Figure \ref{ITEdist2} depicts the distribution of ITE estimates and corresponding ATE associated with the examined outcomes (stroke, bleeding, death and combined). As expected, the use of oral anticoagulants confers protection against stroke and death. On average, patients having anticoagulant are $5.8\%$ ($95\%$ CI $4.0\%$-$7.7\%$) more likely to survive than the control group at the end of 4 quarters (a hazard ratio (HR) equivalent to 0.44), and $14\%$ ($95\%$ CI $9.6\%$-$18.4\%$) more likely to survive at the end of 12 quarters (HR: 0.50). The effect on stroke is about $4\%$ ($95\%$ CI $1.9\%$-$6.1\%$) at 12 quarters (HR: 0.56). These equivalent hazard ratios generally agree with previous meta-analysis\cite{Anonymous:2006is} on the effect of Aspirin, Warfarin and Ximelagatran in patients with non-valvular atrial fibrillation. Treatment effect heterogeneity exists although CATE is always positive ($\mathrm{CI}^*$ increases from  $1.2\%$ to $4.3\%$ from the end of the first quarter to the 12th quarter when the outcome is death). The scale of heterogeneity becomes larger with longer follow-up times, but is accompanied by higher uncertainty, since unmeasured confounding in this framework is more likely to be present with longer follow-up times.

	\begin{table*}[!htbp]%
		\caption{Diagnostic statistics for contributing feature identification \label{CASEImpS}}
		\centering
		\normalsize
		\setlength\tabcolsep{1.3pt}
		\fontsize{8pt}{9pt}\selectfont

		\begin{tabular}{lcccccccccccccccc}
		\toprule
		\textbf{Times indentified} & \multicolumn{4}{c}{\cellcolor[rgb]{ .949,  .949,  .949}\textbf{Adaptive Lasso}} & \multicolumn{4}{c}{\textbf{BART}} & \multicolumn{4}{c}{\cellcolor[rgb]{ .949,  .949,  .949}\textbf{Causal Forest}} & \multicolumn{4}{c}{\textbf{Elastic Net}} \\
		\cmidrule{2-17}  & \cellcolor[rgb]{ .949,  .949,  .949}\textbf{Bld} & \cellcolor[rgb]{ .949,  .949,  .949}\textbf{Dth} & \cellcolor[rgb]{ .949,  .949,  .949}\textbf{Stk} & \cellcolor[rgb]{ .949,  .949,  .949}\textbf{Com} & \textbf{Bld} & \textbf{Dth} & \textbf{Stk} & \textbf{Com} & \cellcolor[rgb]{ .949,  .949,  .949}\textbf{Bld} & \cellcolor[rgb]{ .949,  .949,  .949}\textbf{Dth} & \cellcolor[rgb]{ .949,  .949,  .949}\textbf{Stk} & \cellcolor[rgb]{ .949,  .949,  .949}\textbf{Com} & \textbf{Bld} & \textbf{Dth} & \textbf{Stk} & \textbf{Com} \\
		\midrule
		\rowcolor[rgb]{ .851,  .851,  .851} \textbf{Charlson index} & 30 & 30 &   & 30 & 30 & 30 &   & 30 &   &   &   &   & 30 & 30 &   & 30 \\
		\textbf{Age} & \cellcolor[rgb]{ .949,  .949,  .949} & \cellcolor[rgb]{ .949,  .949,  .949} & \cellcolor[rgb]{ .949,  .949,  .949} & \cellcolor[rgb]{ .949,  .949,  .949} & 30 & 30 & 30 & 30 & \cellcolor[rgb]{ .949,  .949,  .949}30 & \cellcolor[rgb]{ .949,  .949,  .949}30 & \cellcolor[rgb]{ .949,  .949,  .949}30 & \cellcolor[rgb]{ .949,  .949,  .949}30 &   &   &   &  \\
		\rowcolor[rgb]{ .851,  .851,  .851} \textbf{Ticlopidine} &   &   & 30 &   &   &   &   &   & 30 & 30 &   & 30 &   &   &   &  \\
		\textbf{Vascular disease} & \cellcolor[rgb]{ .949,  .949,  .949}29 & \cellcolor[rgb]{ .949,  .949,  .949}30 & \cellcolor[rgb]{ .949,  .949,  .949} & \cellcolor[rgb]{ .949,  .949,  .949} &   &   &   &   & \cellcolor[rgb]{ .949,  .949,  .949} & \cellcolor[rgb]{ .949,  .949,  .949} & \cellcolor[rgb]{ .949,  .949,  .949} & \cellcolor[rgb]{ .949,  .949,  .949} & \multicolumn{1}{l}{29} & \multicolumn{1}{l}{30} &   &  \\
		\rowcolor[rgb]{ .851,  .851,  .851} \textbf{Female} &   &   & 30 &   & 30 &   &   &   &   &   &   &   &   &   & 30 &  \\
		\textbf{Hypertension} & \cellcolor[rgb]{ .949,  .949,  .949} & \cellcolor[rgb]{ .949,  .949,  .949} & \cellcolor[rgb]{ .949,  .949,  .949} & \cellcolor[rgb]{ .949,  .949,  .949}30 &   &   &   &   & \cellcolor[rgb]{ .949,  .949,  .949} & \cellcolor[rgb]{ .949,  .949,  .949} & \cellcolor[rgb]{ .949,  .949,  .949} & \cellcolor[rgb]{ .949,  .949,  .949}30 &   &   &   & \multicolumn{1}{l}{30} \\
		\rowcolor[rgb]{ .851,  .851,  .851} \textbf{Cancer (any malignancy)} &   &   &   &   & 30 & 30 &   & 30 &   &   &   &   &   &   &   &  \\
		\boldmath{}\textbf{$\ldots$}\unboldmath{} & \cellcolor[rgb]{ .949,  .949,  .949} & \cellcolor[rgb]{ .949,  .949,  .949} & \cellcolor[rgb]{ .949,  .949,  .949} & \cellcolor[rgb]{ .949,  .949,  .949} &   &   &   &   & \cellcolor[rgb]{ .949,  .949,  .949} & \cellcolor[rgb]{ .949,  .949,  .949} & \cellcolor[rgb]{ .949,  .949,  .949} & \cellcolor[rgb]{ .949,  .949,  .949} &   &   &   &  \\
		\midrule
		\textbf{Number of Important Covariates} & \cellcolor[rgb]{ .949,  .949,  .949}\textbf{2} & \cellcolor[rgb]{ .949,  .949,  .949}\textbf{7} & \cellcolor[rgb]{ .949,  .949,  .949}\textbf{4} & \cellcolor[rgb]{ .949,  .949,  .949}\textbf{8} & \textbf{7} & \textbf{5} & \textbf{5} & \textbf{5} & \cellcolor[rgb]{ .949,  .949,  .949}\textbf{2} & \cellcolor[rgb]{ .949,  .949,  .949}\textbf{2} & \cellcolor[rgb]{ .949,  .949,  .949}\textbf{1} & \cellcolor[rgb]{ .949,  .949,  .949}\textbf{5} & \textbf{7} & \textbf{4} & \textbf{4} & \textbf{2} \\
		\bottomrule
		\end{tabular}%

	\begin{tablenotes}
	\item Each cell reports the number of times a feature was identified in one of 30 bootstrap random samples of size $8,000$ from $28,972$ estimations (The max value 30 means that the feature was identified in all samples).  The last row reports the number of important covariates identified by each algorithm on each outcome. Here we show the top seven covariates by the total number of times identified across four outcomes and four algorithms. The complete table can be found in Table X of Appendix D. 
	\item Abbreviations: BLD, Bleeding; DTH,  Death; STK, Stroke; COM, Combined Outcomes.
	\end{tablenotes}
	\end{table*} 

	Table \ref{CASEImpS} shows the top seven features controlling ITE.  Age is the only feature contributing to all outcomes. Charlson index and cancer with malignancy at baseline are important variables for HTE on bleeding and death. Hasbled score is important for HTE on bleeding and stroke. When outcomes are combined, important baseline covariates according to BART (which performed best in our simulation scenarios) are: the Charlson index, age, cancer with malignancy, $\mathrm{CHA}_2\mathrm{DS}_2\-\mathrm{VASc}$ Score and a first AF diagnosis in hospital.

	The CATE estimations on four age subgroups (249 observations in $[18,39)$, 2,992 in $[39,60)$, 16,587 in $[60,82)$, and 9,093 in $[82,100]$).  The largest range of CATEs is observed in relation to the probability of death, which over 24 quarters, varies from  $0\%$ to $24.3\%$ for the youngest patients to patients aged 100. This is followed by differences in CATE for stroke that vary from $0\%$ to $10.4\%$). This observed HTE is not statistically significant across the smaller age groups $[18,60)$.  Oral anticoagulants for the group of patients aged 82 and above have a double effect on all outcomes as compared to their effect on patients aged between 60 to 82. Although the standard error grows with time, it does so proportionally to the effect size. Across the groups with age greater than 60 and for death and combined outcomes, the one-side simultaneous confidence intervals ($\mathrm{CI}^*$) are less than 50\% of the value of the corresponding CATE estimation from the second quarter to the 24th quarter.

	CATE variations across the Charlson index for death and combined outcomes (see  Appendix D) exhibit some non-linearity over time. For patients with Charlson index greater than 14, the effect on death grows from $0\%$ to the maximum at the end of the fourth quarter with a difference of $33.8\%$ and dies out to about $11.6\%$ at the 24th quarter. However, this nonlinearity is not observed in those with Charlson index less than 12. It is worth noting that in this manuscript we report the simultaneous confidence interval, which is much conservative than the empirical confidence interval calculated using the  standard deviation. For example, the $95\%$ simultaneous confidence interval of the ATE for death outcome at the end of fourth quarter is $(4.0\%,7.7\%)$, whose corresponding empirical confidence interval is $(5.6\%,6.0\%)$.

\section{Discussion}

	This article presents a framework to estimate heterogeneous treatment effect in terms of survival probabilities.  These outcomes are popular in the evaluation of medical interventions such as patient survival, cancer recurrence, or cardiovascular events, which are subject to loss of follow up.  In contrast, binary data used in machine learning contains a finite number of values where censoring is not allowed. Since machine learning methods have been hugely successful in predicting discrete outcomes,  there has been significant interest in applying them to study causal effect in medical problems, and examples include BART\cite{Hill:2011hv}, Causal Boosting \cite{Hahn:2017tm} and Causal Forest \cite{Wager:2018eq}.  However, these studies have not explained how to estimate HTE on survival outcomes. 

	The enormous challenge in the estimation of causal effects on survival data is to incorporate an estimate of the treatment and censoring mechanism. Existing methods such as Random Survival Forest\cite{Ishwaran:2008gj}, Super-learner \cite{Polley:2011jw} and Deep Neural Networks\cite{Katzman:2018bv,Lee:2018tc,Ren:2018ue}  only account for the censoring mechanism.  Using TMLE  allows the efficient estimation of the average treatment effect as survival probabilities at a single time point by accounting both mechanisms.\cite{benkeser2017improved} As practitioners are often interested in how treatment effect may evolve with time, the study on one-step TMLE demonstrated how the causal effect could be represented by a monotonic survival curve. But it has not been comprehensively examined with simulation experiments or typical clinical data like the ones laid out in this paper, neither has it given a solution to the estimation of effect heterogeneity. 

	An advantage of our proposed framework is that it reports the effect heterogeneity in terms of the difference in survival probabilities under treatment and control conditions. This absolute measure is more informative for clinical decisions compared to relative measures such as the hazard ratio. Secondly, we provide the flexibility of choosing user-specified methods for potential outcome modelling,  important covariates identification and selection bias adjustment. Thirdly, we automate the heterogeneity discovery process to avoid the ad-hoc selection of subgroups. The adoption of bootstrap sampling allows the use of BART when the sample size is large, and computing memory is limited. Fourthly, the adjustment for selection bias was conducted on the identified subgroups to reflect the treatment and censoring mechanisms within the chosen subgroup.  

	The challenge of a good survival outcome model lies in the fact that the sample size decreases with time which makes the estimation later in time highly uncertain. Although Super-learner allows a flexible choice of algorithms for the best classification,  it did not calibrate the model for the optimal probability estimation.  We found that increasing either the sample size or event rate helps to improve the estimation accuracy. On the other hand, the level of treatment confounding had less impact on the accuracy of CATE than the ATE, even though low confounding scenarios generally produce better results ceteris paribus.  Despite the superior results from using BART, our feature discovery process is vulnerable to the estimation of ITE, which has high uncertainty and is unadjusted for selection bias.  We found that both linear and non-linear features with a stronger effect on treatment effect are easier to be identified, but over-shadow the impact of features with less contribution. 

	Several important extensions and refinements are left open. This study explores HTE estimation on typical medical studies where sample sizes are not huge, and there is a moderate amount of meaningful covariates (such as key comorbidities or risk scores).  In future work, we shall explore the usage of the sequential deep learning on high-dimension observational datasets. We recognise that in estimating heterogeneous treatment effects, the structure of the potential outcome model matters more than the strategy for handling selection bias. This is because selection bias is inherently a misspecification problem\cite{Alaa:2018ue}, and hence its impact on non-parametric inference is washed away in a sufficiently large amount of data with overlap.  Nonetheless, in many cases where the sample size is small or a reference class is not representative, we shall properly account for selection bias.  This paper explores the usage of one-step TMLE,  and it would be intriguing to compare it with more recent methods such as the Quasi-Oracle R-learner \cite{2017arXiv171204912N} and the Non-stationary Gaussian Process Regression (NSGP) \cite{Alaa:2018ue}. 

	\section*{Acknowledgements}
We are grateful for helpful feedback from Weixin Cai on the application of one-step Targeted Maximum Likelihood Estimation for time-to-event outcomes. This work was supported by National Health and Medical Research Council, project grant no. 1125414. Ethics to use UK Clinical Practice Research Datalink data was obtained from ISAC (protocol number $17\_093$).




\noindent

\bibliographystyle{unsrt}
\bibliography{AFPaper}%

\end{document}